\documentclass[review,authoryear,12pt]{elsarticle}
\usepackage{lineno}

\usepackage{amsmath,amsfonts,amssymb}
\usepackage{cancel}
\usepackage{caption}
\usepackage{subcaption}
\usepackage{adjustbox}
\usepackage{multirow}
\usepackage{booktabs}
\usepackage[table]{xcolor}
\usepackage{colortbl}
\usepackage{lipsum} 
\usepackage{titletoc}
\usepackage[symbol]{footmisc}

\usepackage{pdflscape}
\usepackage{xurl}
\usepackage{bbding}

\usepackage[ruled,vlined]{algorithm2e}
\usepackage{float}

\usepackage[noabbrev,nameinlink,capitalize]{cleveref}

\newcommand{\yijg}{y_{ijg}}
\newcommand{\gammaig}{\gamma_{ig}}
\newcommand{\deltaig}{\delta_{ig}}
\newcommand{\varepsilonig}{\varepsilon_{ijg}}
\newcommand{\btheta}{\boldsymbol{\theta}}

\usepackage{xspace}

\newcommand{\combat}{ComBat\xspace}
\newcommand{\combatgam}{ComBat-GAM\xspace}

\newcommand{\dcombat}{d-ComBat\xspace}

\newcommand{\fedcombat}{Fed-ComBat\xspace}

\newcommand{\deepcombat}{DeepComBat\xspace}

\newcommand{\Title}{\fedcombat: A Generalized Federated Framework for Batch Effect Harmonization in Collaborative Studies}

\begin{document}

\title{\Title}

\affiliation[1]{organization={Université Côte d'Azur and INRIA, EPIONE team},
            city={Sophia Antipolis},
            country={France}}
\affiliation[2]{organization={University College London, The UCL Hawkes Institute, Department of Medical Physics and Biomedical Engineering},
                city=London,
                country={UK}}
\author[1]{Santiago Silva*\footnote{*Authors with equal contribution\label{first_author}}\corref{cor1}}
\author[1]{Ghiles Reguig*\corref{cor1}}
\author[2]{Neil P. Oxtoby}
\author[2]{Andre Altmann}
\author[1]{Marco Lorenzi}\ead{marco.lorenzi@inria.fr}

\begin{abstract}
The use of multi-centric analyses is crucial for obtaining sufficient sample sizes and representative clinical populations in experimental studies. In this setting, data harmonization techniques are typically employed to address systematic biases and ensure the interoperability of the data. State-of-the-art harmonisation approaches are based on the statistical theory of random effect modeling, allowing to account for either linear of non-linear biases and batch effects. However, optimizing these statistical methods generally requires data centralization at some point during the analysis pipeline, therefore introducing the risk of exposing individual patient information while posing significant data governance issues. 
To overcome this challenge, in this paper we present \fedcombat, a federated framework for batch effect harmonization on decentralized data. \fedcombat enables the preservation of nonlinear covariate effects without requiring centralization of data and without prior parametric hypothesis on the variables to account for.

We demonstrate the effectiveness of \fedcombat against a comprehensive panel of existing approaches based on the state-of-the-art \combat, along with distributed and nonlinear variants. Our experiments are based on extensive simulated data, and on the analysis of multiple cohorts based on 7 neuroimaging studies comprising healthy controls (CI) and subjects with various disorders such as Parkinson’s disease (PD), Alzheimer's disease (AD), and autism spectrum disorder (ASD).

Our results show that in a federated settings, \fedcombat harmonization exhibits comparable results to centralized methods for both linear and nonlinear cases. On real data, harmonized trajectories of the thickness of the right hippocampus across lifespan measured on a set of 7 public studies show comparable results between centralized and federated models and are consistent with the literature when using a nonlinear model. 
\\ The code is publicly available at: \url{https://gitlab.inria.fr/greguig/fedcombat}
\end{abstract}


\maketitle

\section{Introduction}
\label{sec:introduction}
Data harmonization is known to be a crucial factor in tackling data heterogeneity in studies across multiple sites \citep{wachinger2021detect}. Statistical data harmonization aims to address biases and variations among data collected from different sources, ensuring that the data is comparable and can be combined effectively for analysis. Typical harmonization methods, such as \combat, aim at correcting potential biases due to site effects while preserving the ones associated with covariates of interest (e.g., sex, diagnosis, age) \citep{fortin2017harmonization, neurocombat}. The standard formulation of \combat relies on linear mixed effect models to disentangle the site variability from the fixed effects associated with the desired covariates.  More recently, \combatgam extended this framework by reformulating the linear mixed effect functions through generalized additive models (GAMs) to account for nonlinear covariate effects \cite{combatgam}. 
Although \combatgam showed better performance than \combat in the LIFESPAN dataset \citep{combatgam}, it remains limited by design on the parameterization of the non-linear functions, such as polynomial or splines. Moreover, \combatgam requires an explicit definition of covariate interactions, typically limited to additive or multiplicative terms (e.g., $\mathrm{Sex} \times \mathrm{Diagnosis}$). Despite efforts, a gap still exists in the availability of methods for flexible covariate effect preservation without requiring data centralization \cite{gebre2023cross}.

This problem is further exacerbated by the practical setting of real-world multi-centric studies. Due to governance standards and privacy issues, data often cannot be shared among sites in a centralized server. This issue challenges the practical use of standard harmonisation protocols that generally assume data availability. Federated learning (FL) is a popular machine learning paradigm that allows for collaborative model optimization while maintaining data privacy and governance \citep{konevcny2016federated}. In FL, models are trained on data that remains decentralized across multiple institutions or devices, ensuring that sensitive data stays securely within the respective entities \citep{konevcny2016federated}. FL operates by sharing only model updates or parameters, aggregating these updates on a central server, and distributing the updated global model back to each institution. 
\dcombat is a recent extension of \combat to adapt the harmonisation task to the federated setting  \cite{dcombat}. The underlying idea of this approach is to adapt the optimization of \combat in the distributed setting through the decomposition of the data covariance matrix across sites. This operation allows solving the least squares problem associated to the linear \combat function, subsequently allowing the estimation of local sites effects while bypassing the data sharing task.
Nevertheless, \dcombat can be applied only in the linear setting, due to the specific optimization scheme tailored to the ordinary least squares (OLS) problem. The extension of \dcombat to non-linear effect estimation thus requires the investigation of more general distributed optimization schemes compatible with the \combat formulation. This is essential to allow the generalization of the harmonisation task to account for non-linear covariate effects in real-world multi-site settings \cite{bethlehem2022brainNature}.

To fill this methodological gap, in this work we investigate a general formulation of \combat based on flexible and non-parametric models of covariate effects, while allowing for a distributed optimization in a federated learning setup.
Our approach enables a more nuanced representation of covariate effects during the harmonization process while encompassing \combat, \combatgam, and \dcombat as specific cases. 
We benchmarked our proposed methods with existing centralized, distributed, linear and non-linear variants \citep{combat, combatgam, dcombat}. 
The different methods were compared for their ability to harmonize batch effects and preserve the quality of covariate effects on simulated data (\cref{sec:results}). Furthermore, we  performed an evaluation on derived phenotypes from MRI-brain images from seven cohorts, comprising controls and patients with different brain disorders: patients with different subtypes of Parkinson’s disease (PD), Alzheimer's disease (AD) and Autism spectrum disorder (ASD). 
In all our experiments, our proposed models achieved accurate and generally superior harmonization capabilities while maintaining biological covariates information. 
\\ Ultimately, our work extends current possibilities to data harmonization in multi-centric clinical studies to real decentralized setups that are compliant with data governance and sharing constraints. 

\section{Methods}
\label{sec: methods}
\subsection{Generalized \combat model}
\label{sec:generalized-combats}
Following the original \combat formulation proposed by \cite{combat}, let us denote a batch (represented by different scanners protocols, machines, or institutions) indexed by $i \in \{1, 2, ..., S\}$ on a particular phenotype (e.g., a brain region) indexed by $g \in \{1, 2, ..., G\}$. Each batch contains $n_i$ number of observations, and the total number of observations is $N = \sum_i^S n_i$. $S$ can denote for simplicity the number of sites in the study, but it can also be extended to the total number of scanners between sites or any other number of batch effects. We can model a specific phenotype $g$ observed in the $j$-th patient who belongs to the $i$-th site denoted by $\yijg$ as follows:
\begin{align}
    \yijg = \alpha_g + \phi(\mathbf{x}_{ij}, \btheta_g) + \gammaig + \deltaig\varepsilonig , \label{eq:general-combat}
\end{align}

where $\mathbf{x}_{ij}$ denotes the covariate effects expected to be preserved after removing the batch effects (e.g., sex and age), $\alpha_g$ acts as a global fixed intercept (i.e., the mean),  while $\gammaig$ indicates a random intercept that accounts for the site-specific shift. $\varepsilonig$ is a noise model that captures the variability of each phenotype $\varepsilonig \sim \mathcal{N} (0, \sigma_g^2)$, and $\deltaig$ is a multiplicative effect that scales the ``unbiased'' phenotype variability to fit the one at each site.

This formulation generalizes the original linear model proposed by \cite{combat} and the nonlinear covariate effect model proposed by \cite{combatgam}, to account for potentially multivariate and linear fixed functions $\phi(\mathbf{x}; \btheta_g)$ parameterized by $\btheta_g$.  

We introduce the following constraints to the estimation of fixed effect parameters ($\hat{\alpha}_g, \hat{\btheta}_g, \hat{\gamma}_{ig}, \hat{\sigma}_g$):
\begin{align}
    \arg\max_{\hat{\alpha}_g, \hat{\btheta}_g, \hat{\gamma}_{ig}, \hat{\sigma}_g} P(y_{ijg} | \hat{\alpha}_g, \hat{\btheta}_g, \hat{\gamma}_{ig}, \hat{\sigma}_g) \label{eq:mle-max}\\
    \textrm{subject to} \quad \mathbb{E}_g[\hat{\gamma}_{i}]  = \sum_i^S \frac{n_i}{N} \hat{\gamma}_{ig} = 0, \, \forall g \in \{1, ..., G\} \label{eq:constraint1} \\
    \textrm{and} \quad \phi(\mathbf{x}, \btheta_g)\rvert_{x = 0} = 0 \label{eq:constraint2}
\end{align}

A first constraint in \cref{eq:constraint1} is set to allow identifiability of the intercept parameters as explained by \cite{combat}. However, in the same regard, the constraint in \cref{eq:constraint2} ensures that no batch effects are captured by the covariate function $\phi(\cdot)$. This second constraint, despite not being mentioned, is also fulfilled by \cite{combat} and \cite{combatgam}; making \combat and \combatgam a particular case of the proposed formulation in this work.

For a centralized setup, the estimation of all these parameters is performed in three steps: i) maximum likelihood estimation (MLE) for parameters $\hat{\alpha}_g, \hat{\btheta}_g, \hat{\gamma}_{ig}$ (see \cref{eq:mle-max}) and for the phenotype variance $\hat{\sigma}_g^2 = \frac{1}{N} \sum_{ij} \big(\yijg - \hat{\alpha}_g - \phi(\mathbf{x}_{ij}; \hat{\btheta}_g) -\hat{\gamma}_{ig} \big)^2$, ii) residual standardization mapping the residuals to satisfy the form $\yijg \to z_{ijg} \sim \mathcal{N}(\gamma_{ig}, \delta_{ig}^2)$ as follows:
\begin{align}
    z_{ijg} = \frac{ \yijg - \hat{\alpha}_g - \phi(\mathbf{x}_{ij}; \hat{\btheta}_g) }{\hat{\sigma}_g} \label{eq:standardization}
\end{align}

and iii) estimation of the additive and multiplicative batch effects $\hat{\gamma}_{ig}^*$ and $\hat{\delta}_{ig}^*$ as in \cref{eq:eb-priors}, using empirical Bayes (EB) with priors on $\gammaig$ and $\deltaig^2$ to iteratively estimate these parameters as in \cref{eq:gamma-ig-eb} \citep[Sec.~3.2]{combat}.
\begin{align}
    \gamma_{ig} \sim \mathcal{N}(Y_i, \tau_i^2) \quad \textrm{and} \quad \deltaig^2 \sim \textrm{Inverse Gamma}(\lambda_i, \vartheta_i) \label{eq:eb-priors}
\end{align}
\begin{align}
    \gammaig^{*} = \frac{n_i \Bar{\tau}_i^2\hat{\gamma}_{ig} + \deltaig^{2*} \Bar{\gamma}_{ig}}{n_i \Bar{\tau}_i^2 + \deltaig^{2*}},  \, \deltaig^{*} = \frac{\Bar{\theta}_i + \frac{1}{2} \sum_{j=1}^{n_i} (z_{ijg} - \gammaig^*)^2}{\frac{n_i}{2} \Bar{\lambda}_i - 1} \label{eq:gamma-ig-eb}
\end{align}

Lastly, phenotypes can be harmonized while preserving the covariate effects of interest as follows:
\begin{equation}
    \yijg^{\textrm{\tiny\combat}} = \frac{\hat{\sigma}_g}{\hat{\delta}_{ig}^{*}} \big( z_{ijg} - \hat{\gamma}_{ig}^{*} \big) + \hat{\alpha}_g + \phi({\mathbf{x}_{ij}}; \hat{\btheta}_g)
    \label{eq:harmonize}
\end{equation}

In the following section, we will discuss how this formulation facilitates the incorporation of harmonization within the federated learning framework.

\subsection{Federated learning}
\label{sec:fedcombat}
In a standard federated setting we focus on a distributed optimization problem associated with the functional:
\begin{align}
    \arg\min_{\btheta}{F(\btheta)}, \\
    \textrm{where}, \, F(\btheta) := \sum_{i=1}^S \frac{n_i}{N} F_i(\btheta),
    \label{eq:federated-opt-problem}
\end{align}
where $F_i$ is the cost function evaluated at site $i$, $n_i$ is the number of data points available at site $i$, and $N = \sum_i n_i$.
\textit{Federated averaging} (\textsc{FedAVG}) is the standard approach to optimize the problem of Equation (\ref{eq:federated-opt-problem}) \citep{fedavg}, based on the iteration of partial optimization of problem (\ref{eq:federated-opt-problem}) at each site followed by a step of parameters aggregation across clients. In this setting, when local optimization is performed by stochastic gradient descent (SGD), the convergence of the federated optimization process has been demonstrated for both IID and non-IID data distribution across clients \citep{li2019convergence}.

\subsection{SGD-\combat: Flexible and General \combat optimization}

We propose the SGD-\combat model by defining the kernel function $\phi(\mathbf{x}; \btheta_g)$ as a neural network optimized with stochastic gradient descent (SGD). This allows the harmonization process to model covariate effects (defined by the kernel function) using universal approximators such as multi-layer perceptrons (MLP).
\\ Two variants of SGD-\combat are used in our work: a linear one, which we refer to as SGD-\combat Linear, and a non-linear one, modeling the covariate effects with an MLP which we refer to as SGD-\combat MLP. To respect the constraint mentioned in Equation \ref{eq:constraint2}, the layers of the SGD models do not include biases. 
\\ Once the SGD model has been optimized, the parameters $\boldsymbol{\theta} = \{\alpha_g, \btheta_g$, $\gamma_{ig}\}$ are estimated as described in the Equations \ref{eq:gamma-ig-eb}.

\subsection{From SGD-\combat to \fedcombat }

The extension of  SGD-\combat to the federated learning regime follows directly by replacing the global SGD optimization step with an iterative federated scheme, such as \textsc{FedAVG} or \textsc{FedProx} \citep{li2020federated}. In particular, Considering the formulation presented in \cref{eq:general-combat}, the parameters $\boldsymbol{\theta} = \{\alpha_g, \btheta_g$, $\gamma_{ig}\}$ can be optimized as in \cref{eq:federated-opt-problem}, followed by the local estimation of the random effects following the standard \combat routine, for example based on Empirical Bayes (EB) (\cref{eq:gamma-ig-eb,eq:eb-priors}). A description of the steps followed in federated harmonization using \fedcombat is described in \cref{alg:fedcombat} including 
local updates relying on SGD, global updates across sites of shared parameters using \textsc{FedAVG} and random effect estimation.

\begin{algorithm}[!htbp]
\KwData{Non-harmonized phenotypes ($y_{ijg}$) and covariates ($\mathbf{x}_{ij}$).}
\KwResult{Harmonized phenotypes with siloed data.}
$\mathbf{x} \gets$ \textsc{FederatedStandardization($\mathbf{x}$})\;
\BlankLine
Estimation of fixed effects and random intercept:\\
initialize of parameter space $\boldsymbol{\Omega} := \{ \boldsymbol{\theta}_g, \alpha_g, \gamma_{ig}\}_{g=1}^{G}$\;
\While{not converged}{ 
    \ForEach{site $i$}{
        $\boldsymbol{\Omega}_i^{t} \gets \boldsymbol{\Omega}^{(t+1)}$ \tcp*[r]{Initialization.}
        \tcp{Partial local optimization using SGD.}
        \ForEach{local gradient step $t$}{
            $\boldsymbol{\Omega}_i^{(t+1)} = \boldsymbol{\Omega}_i^{(t)} - \eta \nabla_{\boldsymbol{\Omega_i}} F(\boldsymbol{\Omega_i}^{(t)})$ \;
        }
    }
    \tcp{Aggregate and update every parameter using FedAVG.}
    $\boldsymbol{\Omega}^{(t+1)} \gets \sum_i \frac{n_i}{N} \boldsymbol{\Omega}_i^t$
}

\ForAll{site $i$}{
    \tcp{Standardize using estimated parameters}
    $z_{ijg} \gets \frac{ \yijg - \hat{\alpha}_g - \phi(\mathbf{x}_{ij}; \hat{\btheta}_g) }{\hat{\sigma}_g}$

    \tcp{Estimate $\gamma_{ig}^*$ and $\delta_{ig}^*$ using EB}
    $\gamma_{ig}^*, \delta_{ig}^* \gets$ \textsc{EmpiricalBayes($\mathbf{x}$}) \tcp*[r]{\cref{eq:eb-priors,eq:gamma-ig-eb}}

    \tcp{Correct data}
    \KwRet $\yijg^{\textrm{\tiny\combat}} \gets \frac{\hat{\sigma}_g}{\hat{\delta}_{ig}^{*}} \big( z_{ijg} - \hat{\gamma}_{ig}^{*} \big) + \hat{\alpha}_g + \phi({\mathbf{x}_{ij}}; \hat{\btheta}_g)$
}
\caption{\fedcombat}
\label{alg:fedcombat}
\end{algorithm}


\section{Materials}
We evaluated SGD-\combat and \fedcombat on a synthetic benchmark accounting for different sources of batch- and covariate-wise heterogeneity as well as different data/covariates generation schemes. In addition we benchmarked this approach on a collection of seven cohorts corresponding to different studies in neurodegenerative disorders including participants diagnosed with or at risk of developing Parkinson’s disease, Alzheimer’s disease and Autism spectrum disorder.

\subsection{Synthetic data}
\label{sec:materials-synthetic-data}
Following the previous formulation of Equation (\ref{eq:general-combat}), data was generated as such:
\[ \yijg = \alpha_g + \phi(\mathbf{x}_{ij}, \btheta_g) + \gammaig + \deltaig\varepsilonig \]

with 
\begin{itemize}
    \item $\gamma_i \sim \mathrm{Uniform}(-5, 5)$.
    \item $\delta_{ig}\sim \mathrm{Uniform}(2, 5)$ and $\varepsilonig \sim \mathrm{Uniform}(1, 4)$.
    \item $\phi$ is a kernel function with parameters $\btheta_g\sim\mathrm{Uniform}(-10, 10)$ allowing to modulate the dependence of the phenotype $\yijg$ with the covariates $\mathbf{x}_{ij} \sim \mathcal{N}(0; 1)$
\end{itemize}

\subsection{Brain MRI-Data}
We evaluated our method on a cross-sectional cohort of $13479$ participants from seven public studies, utilizing MRI-derived phenotypes from baseline structural T1-weighted magnetic resonance imaging (MRI).

\begin{figure}[!hbtp]
        \centering
        \includegraphics[width=\textwidth]{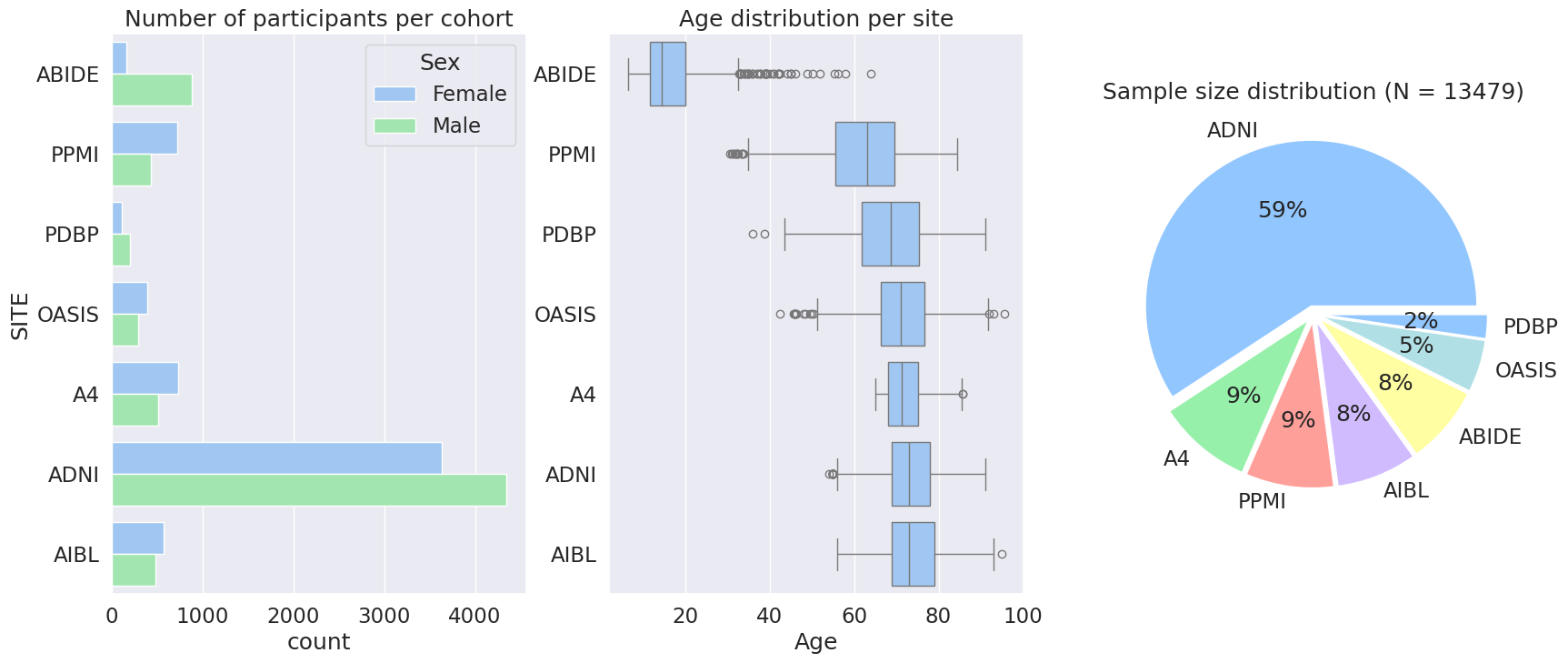}
        \caption{Population pyramid representing the demographics of the real data used in this study. The left panel shows the distribution of sex, the middle panel shows the distribution of age, and the right panel shows the sample size distribution. Cohorts were sorted by ascending median age. The demographics for the population are described in \cref{tab:demographics}.}
        \label{fig:population-pyramid-real}
\end{figure}

Detailed demographic information can be found in Table~\ref{tab:demographics}. Covariate distribution across the multiple cohorts used in this work can be observed in \cref{fig:population-pyramid-real}, clearly denoting the non-IID nature of this dataset.

\subsubsection{Image processing}
For all datasets, with the exception of ABIDE I, cortical thickness and subcortical volume were extracted from MRI scans using FreeSurfer v7.1.1 (documented and freely available for download online at: \url{http://surfer.nmr.mgh.harvard.edu/}) \citep{dale1999cortical,fischl2004automatically}. Steps for phenotypical measures extraction included skull stripping \citep{ashburner2005unified}, Non-uniform intensity Normalization (N3) \citep{sled1998nonparametric}, segmentation using the Desikan–Killiany atlas \citep{desikan2006automated} and extraction of the MRI-derived phenotypes.
\\ The ABIDE I phenotypes were downloaded from the related sources\footnote{\url{https://github.com/dfsp-spirit/abide_preproc_smri_freesurfer6}} and were extracted using FreeSurfer v6. We chose to include the ABIDE data in our analysis to experiment with additional batch effects and variability, and to emphasize the non linearity induced by the differences in brain development patterns between young and elder populations. 

Harmonization was carried out by preserving age, sex, diagnosis and intracranial volume (ICV / eTIV), and each study was considered a site whose systematic biases were expected to be corrected \citep{reynolds2022combat}.

\begin{table}[H]
\caption{Subject demographics of real data used in this work.}
\label{tab:demographics}
\begin{adjustbox}{max width=\textwidth}
\begin{tabular}{ccrrll}
\multicolumn{1}{l}{}                                        & \multicolumn{1}{l}{}                                      & \multicolumn{2}{c}{\cellcolor[HTML]{BDBDBD}\textbf{N}}                                                                  & \multicolumn{2}{c}{\cellcolor[HTML]{BDBDBD}\textbf{Age in years $\pm$ SD  [age range]}}                                 \\
\rowcolor[HTML]{D4D4D4} 
\cellcolor[HTML]{BDBDBD}\textbf{Site}                       & \cellcolor[HTML]{BDBDBD}\textbf{Group  \textbackslash{} Sex} & \multicolumn{1}{c}{\cellcolor[HTML]{D4D4D4}\textbf{Female}} & \multicolumn{1}{c}{\cellcolor[HTML]{D4D4D4}\textbf{Male}} & \multicolumn{1}{c}{\cellcolor[HTML]{D4D4D4}\textbf{Female}} & \multicolumn{1}{c}{\cellcolor[HTML]{D4D4D4}\textbf{Male}} \\
\rowcolor[HTML]{FFFFFF} 
\cellcolor[HTML]{D4D4D4}\textbf{A4} & \cellcolor[HTML]{D4D4D4}\textbf{CU} & 731 & 515 & 71.3 $\pm$ 4.5 \hfill [65.0 - 85.7] & 72.9 $\pm$ 5.1 \hfill [65.0 - 85.7] \\
\rowcolor[HTML]{F3F3F3} 
\cellcolor[HTML]{D4D4D4}  & \cellcolor[HTML]{D4D4D4}\textbf{Autism} & 62 & 443 & 16.3 $\pm$ 7.9 \hfill [8.1 - 45.0]  & 17.2 $\pm$ 8.6 \hfill [7.0 - 64.0] \\
\rowcolor[HTML]{FFFFFF} 
\multirow{-2}{*}{\cellcolor[HTML]{D4D4D4}\textbf{ABIDE}}  & \cellcolor[HTML]{D4D4D4}\textbf{CU} & 95 & 435 & 14.8 $\pm$ 5.5 \hfill [7.8 - 32.0] & 17.2 $\pm$ 7.8 \hfill [6.5 - 56.0] \\
\rowcolor[HTML]{F3F3F3} 
\cellcolor[HTML]{D4D4D4} & \cellcolor[HTML]{D4D4D4}\textbf{AD} & 710 & 891 & 73.1 $\pm$ 7.8 \hfill [55.0 - 91.0] & 74.5 $\pm$ 7.0 \hfill [55.0 - 90.0] \\
\rowcolor[HTML]{FFFFFF} 
\cellcolor[HTML]{D4D4D4} & \cellcolor[HTML]{D4D4D4}\textbf{CU} & 1477 & 1327 & 72.8 $\pm$ 6.0 \hfill [55.0 - 90.0] & 73.9 $\pm$ 6.2 \hfill [60.0 - 90.0] \\
\rowcolor[HTML]{FFFFFF} 
\multirow{-3}{*}{\cellcolor[HTML]{D4D4D4}\textbf{ADNI}} & \cellcolor[HTML]{D4D4D4}\textbf{MCI}& 1451 & 2128 & 71.5 $\pm$ 7.7 \hfill [55.0 - 88.0] & 73.6 $\pm$ 7.0 \hfill [54.0 - 90.0] \\
\rowcolor[HTML]{FFFFFF} 
\cellcolor[HTML]{D4D4D4}  & \cellcolor[HTML]{D4D4D4}\textbf{AD} & 60 & 55 & 75.4 $\pm$ 7.4 \hfill [58.0 - 93.0] & 76.0 $\pm$ 7.6 \hfill [60.0 - 89.0]\\
\rowcolor[HTML]{F3F3F3} 
\cellcolor[HTML]{D4D4D4} & \cellcolor[HTML]{D4D4D4}\textbf{CU} & 445 & 345 & 72.8 $\pm$ 6.5 \hfill [60.0 - 91.0] & 73.6 $\pm$ 6.6 \hfill [60.0 - 92.0] \\
\rowcolor[HTML]{FFFFFF} 
\multirow{-3}{*}{\cellcolor[HTML]{D4D4D4}\textbf{AIBL}} & \cellcolor[HTML]{D4D4D4}\textbf{MCI} & 67 & 85 & 76.0 $\pm$ 7.8 \hfill [56.0 - 95.0] & 75.0 $\pm$ 6.0 \hfill [64.0 - 88.0] \\
\rowcolor[HTML]{FFFFFF} 
\cellcolor[HTML]{D4D4D4} & \cellcolor[HTML]{D4D4D4}\textbf{AD} & 59 & 76 & 76.1 $\pm$ 6.9 \hfill [62.9 - 95.6] & 76.5 $\pm$ 6.9 \hfill [62.9 - 95.6] \\
\rowcolor[HTML]{FFFFFF} 
\multirow{-2}{*}{\cellcolor[HTML]{D4D4D4}\textbf{OASIS}}    & \cellcolor[HTML]{D4D4D4}\textbf{CU} & 333 & 218 & 68.8 $\pm$ 8.7 \hfill [45.6 - 93.1] & 70.7 $\pm$ 8.3 \hfill [42.5 - 91.9] \\
\rowcolor[HTML]{F3F3F3} 
\cellcolor[HTML]{D4D4D4}\textbf{PDPB} & \cellcolor[HTML]{D4D4D4}\textbf{PD} & 75 & 120  & 69.4 $\pm$ 8.2 \hfill [50.8 - 90.0] & 68.1 $\pm$ 10.6 \hfill [38.8 - 91.0] \\
\rowcolor[HTML]{FFFFFF} 
\cellcolor[HTML]{D4D4D4} & \cellcolor[HTML] {D4D4D4}\textbf{CU} & 34 & 42 & 66.0 $\pm$ 10.1 \hfill [50.7 - 84.4] & 64.7 $\pm$ 11.2 \hfill [36.0 - 84.3] \\
\rowcolor[HTML]{F3F3F3}
\cellcolor[HTML]{D4D4D4} & \cellcolor[HTML] {D4D4D4}\textbf{LBD} & 4 & 38 & 74.0 $\pm$ 6.9 \hfill [67.9 - 81.9] & 68.1 $\pm$ 8.0 \hfill [45.1 - 82.7] \\
\rowcolor[HTML]{F3F3F3} \cellcolor[HTML]{D4D4D4} & \cellcolor[HTML]{D4D4D4}\textbf{GenCohort PD}& 45 & 38 & 65.4 $\pm$ 8.5 \hfill [32.2 - 78.6] & 65.5 $\pm$ 8.0 \hfill [51.7 - 81.2]\\
\rowcolor[HTML]{FFFFFF} 
\cellcolor[HTML]{D4D4D4} & \cellcolor[HTML]{D4D4D4}\textbf{GenCohort Unaff} & 47 & 73 & 61.1 $\pm$ 7.9 \hfill [48.8 - 78.3] & 61.2 $\pm$ 8.4 \hfill [33.7 - 84.3]\\
\rowcolor[HTML]{F3F3F3} 
\cellcolor[HTML]{D4D4D4} & \cellcolor[HTML]{D4D4D4}\textbf{PD} & 415 & 207 & 62.5 $\pm$ 9.8 \hfill [34.8 - 84.2] & 61.3 $\pm$ 9.5 \hfill [33.5 - 81.7] \\
\rowcolor[HTML]{FFFFFF} 
\cellcolor[HTML]{D4D4D4} & \cellcolor[HTML]{D4D4D4}\textbf{Prodromal} & 32 & 5 & 69.7 $\pm$ 6.3 \hfill [60.8 - 84.0] & 70.7 $\pm$ 2.7 \hfill [67.1 - 73.2] \\
\rowcolor[HTML]{F3F3F3} 
\multirow{-6}{*}{\cellcolor[HTML]{D4D4D4}\textbf{PPMI}}     & \cellcolor[HTML]{D4D4D4}\textbf{SWEDD} & 54 & 28 & 63.6 $\pm$ 9.4 \hfill [39.3 - 78.9] & 60.3 $\pm$ 10.1 \hfill [39.5 - 79.6] \\               
\rowcolor[HTML]{FFFFFF} 
\multirow{-6}{*}{\cellcolor[HTML]{D4D4D4}} & \cellcolor[HTML]{D4D4D4}\textbf{CU} & 130 & 84 & 60.7 $\pm$ 11.8 \hfill [30.6 - 82.8] & 58.9 $\pm$ 10.6 \hfill [31.0 - 81.9] \\
\end{tabular}
\end{adjustbox}
\end{table}

\section{Results}
\label{sec:results}

Two versions of \fedcombat were used: the first one sets the kernel as a linear model (\fedcombat Linear) and the second one sets $\phi$ as a multi-layer perceptron (\fedcombat MLP). These two models were compared with their centralized versions optimized with gradient descent, which we refer to as SGD-ComBat Linear for the linear model and SGD-ComBat MLP for the non linear one.
\\ \combat\footnote{\url{https://github.com/Jfortin1/neuroCombat}} (a.k.a. NeuroComBat), \combatgam\footnote{\url{https://github.com/rpomponio/neuroHarmonize}} as well as the distributed version \dcombat\footnote{\url{https://github.com/andy1764/Distributed-ComBat}} were used for comparison. The characteristics of the models are reported in Table \ref{tab:methods}. All implementations were done in Python.

\begin{table}
    \adjustbox{max width=\textwidth}{
    \begin{tabular}{lrr}
        \rowcolor[HTML]{BDBDBD} 
        {\color[HTML]{1A1A1A} \textbf{Model}} & \multicolumn{1}{c}{\cellcolor[HTML]{BDBDBD}{\color[HTML]{1A1A1A} \textbf{Non Linear}}} & \multicolumn{1}{c}{\cellcolor[HTML]{BDBDBD}{\color[HTML]{1A1A1A} \textbf{Distributed}}}\\
        \rowcolor[HTML]{FFFFFF} 
        
        {\color[HTML]{1A1A1A} NeuroComBat} & \color[HTML]{1A1A1A} \XSolidBrush & \XSolidBrush \\
        \rowcolor[HTML]{F3F3F3} 
        {\color[HTML]{1A1A1A} SGD-ComBat Linear} & \color[HTML]{1A1A1A} \XSolidBrush & \XSolidBrush \\
        \rowcolor[HTML]{FFFFFF} 
        {\color[HTML]{1A1A1A} \dcombat} & \color[HTML]{1A1A1A} \XSolidBrush & \Checkmark \\
        \rowcolor[HTML]{F3F3F3} 
        {\color[HTML]{1A1A1A} \fedcombat Linear} & \color[HTML]{1A1A1A} \XSolidBrush & \Checkmark \\
        \rowcolor[HTML]{FFFFFF}
        {\color[HTML]{1A1A1A}\combatgam} & \color[HTML]{1A1A1A} \Checkmark & \XSolidBrush \\
        \rowcolor[HTML]{F3F3F3}
        {\color[HTML]{1A1A1A} SGD-ComBat MLP} & \color[HTML]{1A1A1A} \Checkmark & \XSolidBrush \\
        \rowcolor[HTML]{FFFFFF}
        {\color[HTML]{1A1A1A} \fedcombat MLP} & \color[HTML]{1A1A1A} \Checkmark & \Checkmark \\
    \end{tabular}}
    \caption{Methods used in our experiments and their characteristics}
    \label{tab:methods}
\end{table}

\subsection{Synthetic data}
\label{sec:results-synthetic-data}

In this experiment, the covariates $\mathbf{x}_{ij}$ were simulated in two dimensions and three strategies were used in the generation of the unbiased phenotype $\phi(\mathbf{x}_{ij}, \btheta_g)$: 
\begin{itemize}
    \item linear generation: the kernel is a simple linear model \[\phi_{linear}(\mathbf{x}; \boldsymbol{\theta}) = \mathbf{x}^\intercal \boldsymbol{\theta}\].
    \item nonlinear generation: the kernel is a quadratic model \[\phi_{quadratic}(\mathbf{x}; \boldsymbol{\theta}) = \mathbf{x}^\intercal \boldsymbol{\theta} +  (\mathbf{x} \odot \mathbf{x})^{\intercal}\boldsymbol{\theta}\] 
    where $\odot$ is the Hadamard product. 
    \item mixed generation: the kernel is a mix of a linear model and a quadratic model \[\phi_{mixed}(\mathbf{x}; \boldsymbol{\theta}) = \phi_{linear}({x_1}; \boldsymbol{\theta_1}) +  \phi_{quadratic}({x_2}; \boldsymbol{\theta_2})\]  
\end{itemize}
The number of clients/sites was set to $10$ and the simulation was repeated 50 times.
The MLP is composed of two hidden layers of $50$ neurons, a first one followed by a ReLU activation and a second one followed by a hyperbolic tangent, and a final output layer. The learning rate was set at $10^{-3}$ with $300$ epochs for the centralized models (SGD-ComBat Linear and SGD-ComBat MLP) and $100$ rounds of $3$ epochs for the federated models.
\\ The residuals were computed as

\[ residuals = \yijg^{\textrm{\tiny\combat}} - \phi({\mathbf{x}_{ij}}; \hat{\btheta}_g) - \hat{\alpha}_g = \frac{\hat{\sigma}_g}{\hat{\delta}_{ig}^{*}} \big( z_{ijg} - \hat{\gamma}_{ig}^{*} \big). \]

\begin{figure}[!hbt]
    \centering
    \includegraphics[width=\textwidth]{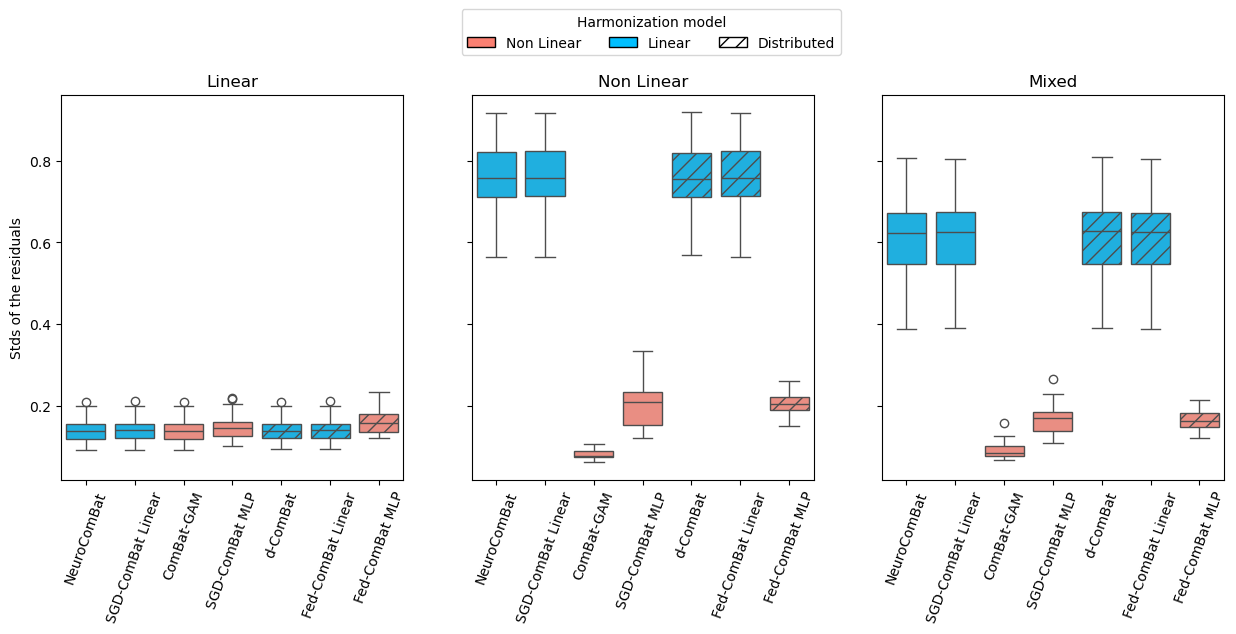}
    \caption{Boxplot of the means of the residuals over 50 runs of simulation. The left, center and right plots respectively show the results for linear, non-linear and mixed simulated data. We note that federated models achieve the similar results as their centralized counterparts, for both linear and nonlinear data. }
    \label{fig:boxplot-residuals}
\end{figure}

\begin{figure}[htbp]
        \hspace{-3.0cm}
        \includegraphics[width=1.4\textwidth]{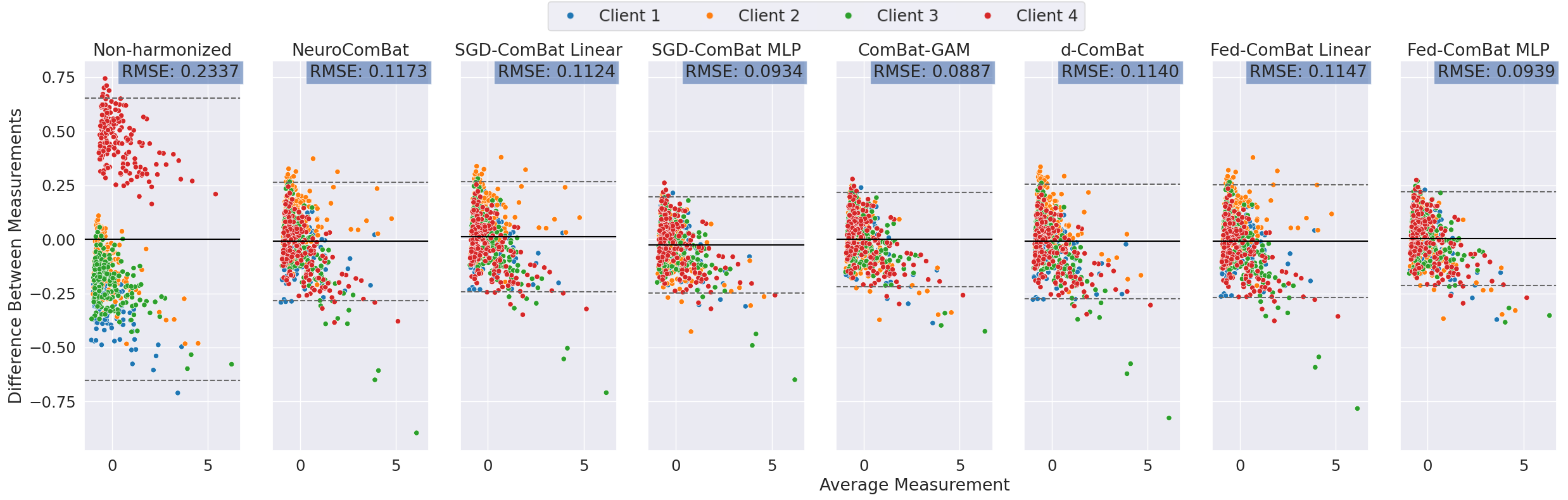}
        \caption{Bland-Altman plots and root mean square errors (RMSE) contrasting the different centralized and federated harmonization methods against the groundtruth. We note that the linear models all show higher RMSE compared to the nonlinear ones. The nonlinear models also show more homogeneous data distributions.}
        \label{fig:bland_altman}
\end{figure}

The standard deviations of the residuals of each of the $50$ runs were computed for each model. Their distributions are reported on Figure \ref{fig:boxplot-residuals} for linear, non linear and mixed simulated data. We observe, as expected, that linear models (in blue) show higher standard deviation residuals in the presence of non linear data whether it is from the non linear or mixed simulation. Regarding the non linear models, \combatgam yields the lowest standard deviations of the residuals and both our proposed non linear methods (SGD \combat MLP and Fed-\combat MLP) obtain similar results which are really close to the \combatgam ones. 
\\ Figure \ref{fig:bland_altman} shows the Bland-Altmann plot for a given non-linear simulation. The Figure shows a clear distinction of data heterogeneity between linear and nonlinear models. We also observe that the federated models achieve similar results to the centralized ones both visually and in RMSE values. 

\subsection{Brain MRI-data}
\label{sec:results-brain-data}
\begin{figure}[htbp]
        \centering
        \includegraphics[width=0.65\textwidth]{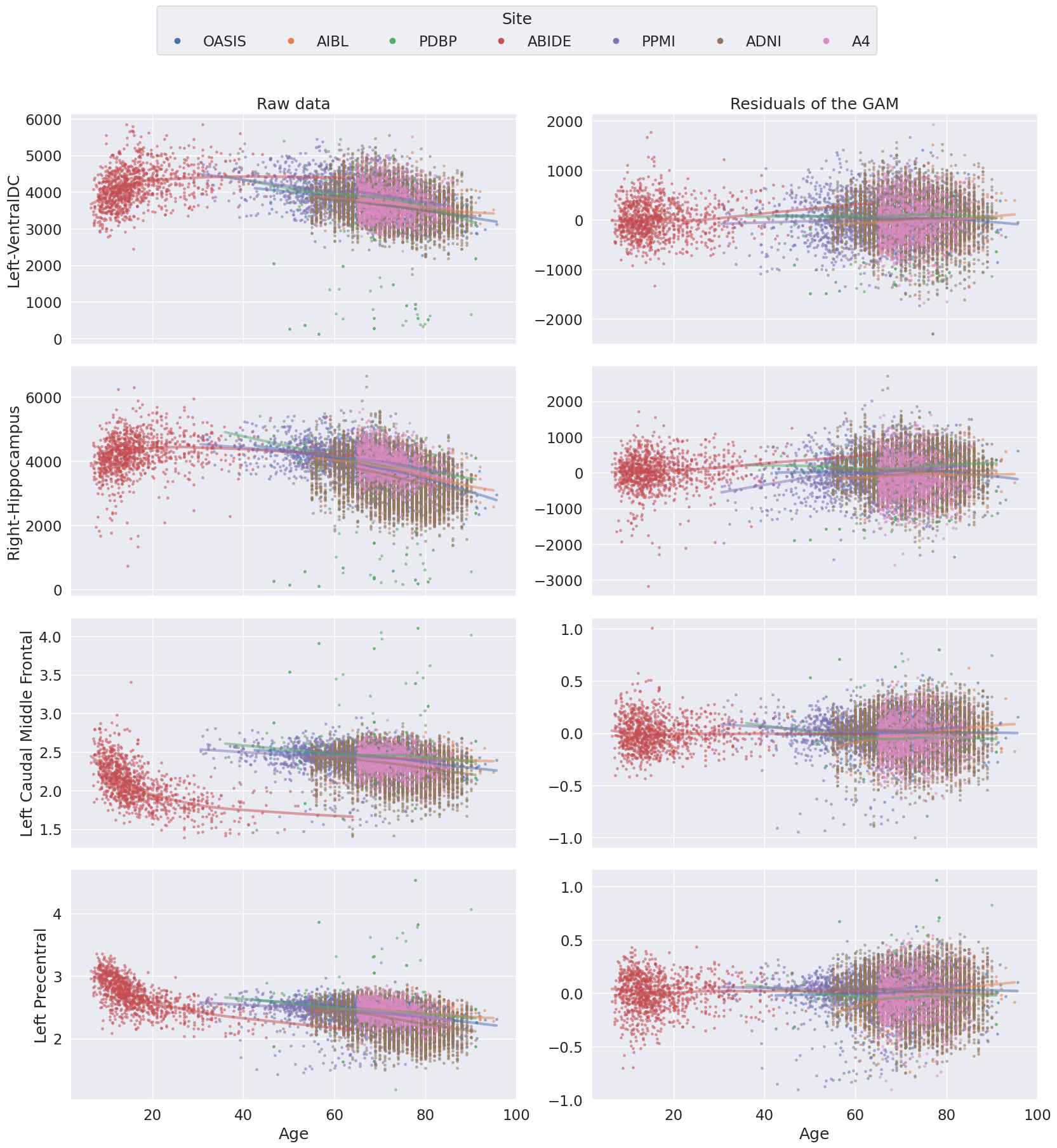}
        \caption{Results of regression using a GAM on the top 4 regions with the most discrepant AIC values on the left and residual plots illustrating goodness of fit of a GAM on the right. From the top to the bottom row: 1) Left ventral diencephalon; 2) Right hippocampus; 3) Left caudal middle frontal gyrus; 4) Left precentral gyrus. On the left, we observe nonlinearity of the trajectories with different rates of atrophy across lifespan. On the right, we see the residuals of the fitted GAM which are centered towards 0 across lifespan, showing that a GAM suitably learns the nonlinearity of the trajectories} 
        \label{fig:nonlinear-evidence-regresion}
\end{figure}

To provide evidence for nonlinear covariate effects of age on brain phenotypes, we compared the goodness of fit of two models: a linear model and a generalized additive model (GAM). Data was first centralized and harmonized accounting for study as the batch source using \combat. The criterion used to evaluate the presence of nonlinearities was the difference of the Akaike Information Criterion (AIC) between the GAM and the linear model, where a negative difference indicates the improved fit of the non-linear GAM. \cref{fig:nonlinear-evidence-regresion} shows the AIC metric across the regions of the brain using the Desikan-Killiany parcellation and cortical thickness across regions and the ASEG atlas for subcortical volumes as the dependent variables. Data was controlled for sex, diagnosis, ICV, and diagnosis group, and age was considered an independent variable. \Cref{fig:nonlinear-evidence-regresion} shows the top four regions that are better explained by a nonlinear model of age in terms of AIC, as well as the residuals after the regression illustrating remaining effects or trends.

\begin{figure}[htbp]
    \vspace{-4.cm}
    
        \includegraphics[width=\textwidth]{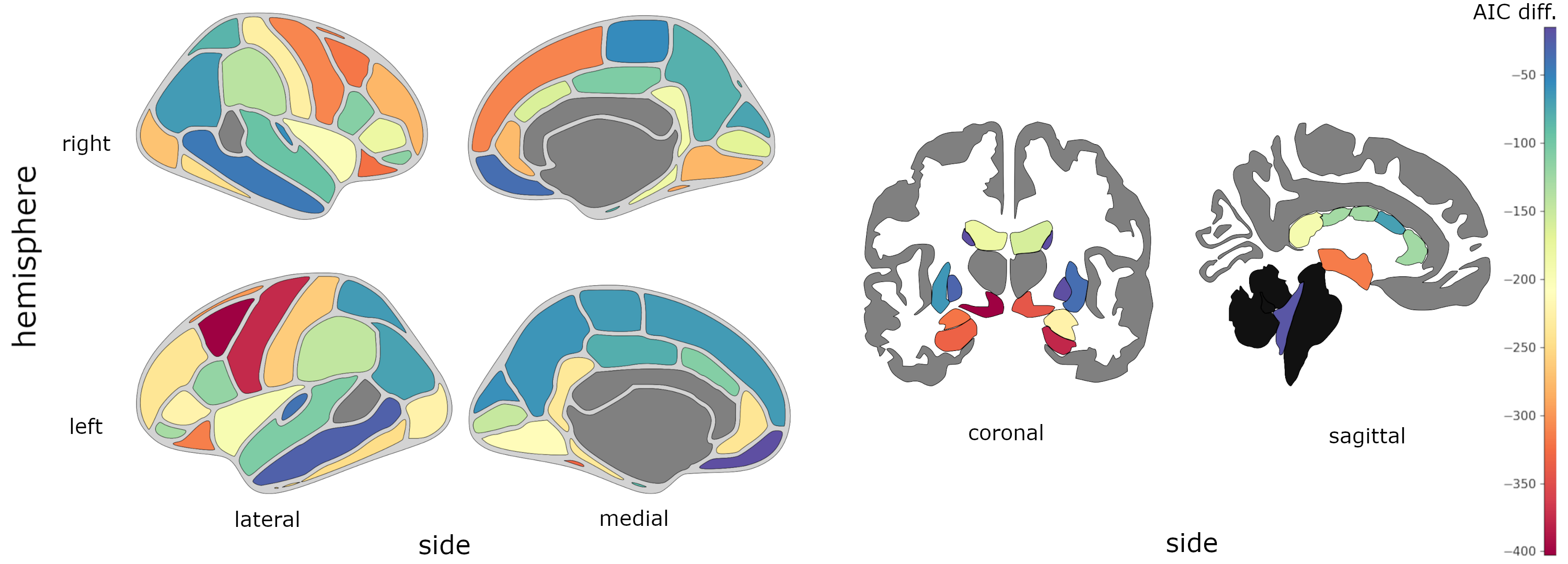}
        \caption{Brain regions where a generalized additive model (GAM) provides a better fit than a linear model, as measured by the difference in Akaike Information Criterion (AIC) between the two models. The parameters for the GAM are set as in \cite{combatgam}, with age as a smoothing term, B-splines with 10 degrees of freedom or control points uniformly distributed between the minimum and maximum values, and a maximum polynomial of 3rd degree. A negative difference indicates a better fit with the GAM, while a positive difference indicates a better fit with the linear model. }
        \label{fig:aic-results}
\end{figure}

Then data was decentralized and we illustrate the age trajectories by regrouping participants in two trajectories: cognitively unimpaired were labeled as ``Healthy controls`` while those diagnosed with a particular neurological disorder were labeled as ``Atypical''. The harmonized phenotypes were obtained using Neuro\combat as linear centralized method, \combatgam as nonlinear centralized, \dcombat as linear federated method, our adaptation \fedcombat using a linear model or an MLP, and finally \fedcombat using the same MLP architecture as described previously (two hidden layers of $50$ neurons respectively followed by a ReLU activation and a hyperbolic tangent, and a final output layer). 

\begin{figure}[!hbt]
    \hspace{-2.cm}
    \includegraphics[width=1.2\textwidth]{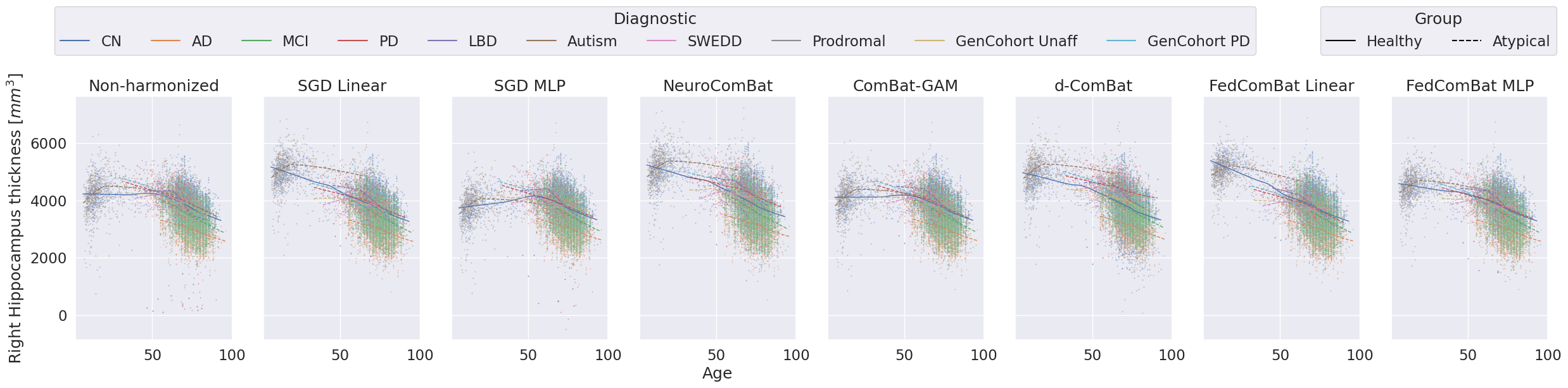}
    \caption{Comparison of age trajectories of the right hippocampus. The columns represent conditions under which the data is shown: non-harmonized (first column), and harmonized using SGD-ComBat Linear (centralized linear SGD based model),  SGD-ComBat MLP (centralized using MLP), Neuro\combat, \combatgam, \dcombat, \fedcombat Linear (federated equiv. to \dcombat but SGD based) and \fedcombat MLP (decentralized). Trajectories were adjusted for sex and ICV. We observe that harmonization using linear models (SGD Linear, Neuro\combat and \fedcombat Linear) show atrophy patterns that are not consistent with the literature while the nonlinear ones (SGD MLP, \combatgam and \fedcombat MLP) better match the different rates of atrophy across lifespan regarding controls (CN).}
    \label{fig:trajectories-real}
\end{figure}

After estimation of each harmonisation method, we assessed the resulting age trajectories by fitting a GAM on the harmonised data by controlling for Sex and ICV. Resulting trajectories are illustrated in \cref{fig:trajectories-real}. The results show that all linear models yield similar trajectories with a continuous decrease throughout lifespan, shown by uniform slope in the trajectory. Regarding the nonlinear models, our proposed \fedcombat MLP shows a slight decrease in the beginning followed by a steeper drop starting around 40/50, while our centralized proposed model SGD MLP shows a slow increase until 50/60 before dropping. Finally, the \combatgam model shows a steadier curve until 50/60 before decreasing. Even though the nonlinear trajectories seem conflicting, they also reflect the non consensus of the evolution of cortical thickness present in the litterature where some works show a slow decrease since early age \citep{ducharme2016trajectories} and others show an inverted U-shape evolution \citep{coupe2017towards}. 

\begin{figure}[!hbpt]
    \centering
    \includegraphics[width=\textwidth]{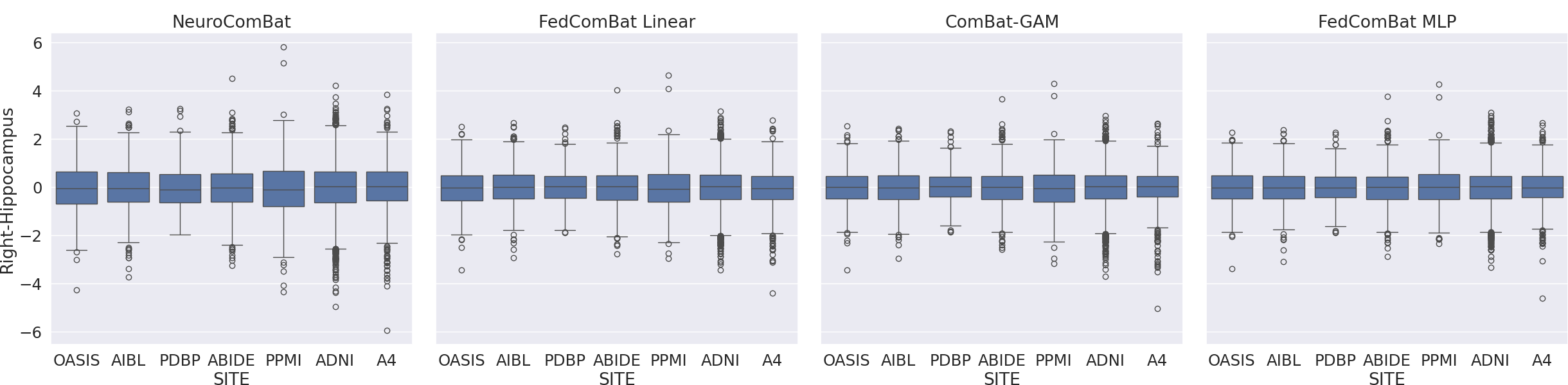}
    \caption{Residual terms after harmonization of the thickness of the right hippocampus for federated methods and their centralized equivalent from the literature. While for each method the distribution of the residuals is centered around 0, the residual variance across dataset is significantly reduced for \fedcombat and \combatgam as compared to Neuro\combat ($p<0.05$ Bartlett test).}
    \label{fig:residuals-sel}
\end{figure}

\cref{fig:residuals-sel} shows the residuals of each distributed model comprehended in this work after harmonization. The residual harmonized term $\frac{\hat{\sigma}_g}{\hat{\delta}_{ig}^*} \left( z_{ijg} - \hat{\gamma}_{ig}^* \right)$ should follow a normal standard noise distribution by definition and should not have center residual effects. \cite{combatgam} proposed fitting a GAM and then extract the residuals. However, by definition, fitting a GAM will work better with a \combat model that is GAM-based. Instead, we use the corresponding model for each method (e.g., linear for \combat, GAM for \combatgam and MLP for \fedcombat MLP). While for each method the distribution of the residuals is centered around 0, the residual variance across dataset is significantly reduced for \fedcombat and \combatgam as compared to Neuro\combat ($p<0.05$ Bartlett test).

\section{Discussion and Conclusion}
\label{sec:discussion_conclusion}

In this paper, we presented \fedcombat, a generalized approach for data harmonization using federated learning. Specifically designed for scenarios where data sharing is restricted and nonlinear covariate effects might be present, \fedcombat provides an extension to existing centralized linear methods such as Neuro\combat. It incorporates the advantages of these methods without necessitating data centralization or explicit specification of nonlinearities and potential interactions.

The performance of \fedcombat, both in terms of its linear and multi-layer perceptron (MLP) versions, is comparable to that of centralized methods. This was evident from the similarity in Bland-Altman plots, despite subtle differences due to the inherent variability in federated setups compared to centralized ones. The MLP version of \fedcombat efficiently captured nonlinearities, offering a more generalizable solution than \combatgam.

Moreover, \fedcombat does not require explicit decomposition of covariates or specification of interactions between them, unlike \combatgam. This flexibility is facilitated by the capacity of complex approximating functions, such as MLPs, to capture nonlinearities and interactions due to its multivariate nature. The general formulation of the \combat method that we proposed allows the use of a wider variety of models to better suit the specificities of each harmonization problem, as we have shown using an MLP in our experiments.  

While this work showed the effectiveness of \fedcombat in handling complex harmonization problems, future research directions could involve evaluating \fedcombat's performance on larger cohorts, such as in \cite{bethlehem2022brainNature}. This would provide additional evidence of \fedcombat's effectiveness in addressing harmonization problems, particularly in scenarios where nonlinear effects are prevalent as was shown for the evolution of several brain regions throughout the entire lifespan by \cite{coupe2017towards}.

\section{Limitations and Future Directions}
\label{sec:limitations_challenges_future}

\fedcombat has shown to be a promising tool for data harmonization in federated learning settings, but it has some inherent limitations that suggest avenues for future exploration. Besides the exploration of new or improved models, required work to make this method available in platforms like COINSTAC \citep{plis2016coinstac} and Fed-BioMed \citep{silva2020fed} will allow real-scenario applications where our approach could be properly evaluated.

Another aspect to consider is the heterogeneity across participating institutions, which could impact the performance of \fedcombat, and of \combat-based correction methods in general. Extreme disparities in data distributions, demographic compositions, or data collection protocols might challenge the method's ability to rectify biases efficiently. This indicates the need to appraise the applicability of \fedcombat or any \combat method in such situations carefully. Furthermore, the current implementation employs Stochastic Gradient Descent (SGD) for optimization which can be improved, especially when dealing with such heterogeneous data. To overcome this, future work could explore methods like FedProx \citep{li2020federated} that provide an efficient and robust solution to SGD in the presence of data heterogeneity.

Beyond these limitations inherent to \fedcombat, there are challenges associated with federated learning in general, such as the requirement for significant computational resources and stable connectivity at each site. Acknowledging these, future research could focus on enhancing the robustness of \fedcombat against such practical constraints.

The current implementation of \fedcombat focuses on derived phenotypes, mirroring the first version of \combat. However, it would be worthwhile to investigate the method's direct application to imaging data in the future. By modifying $\phi(\mathbf{x}_{ij}, \boldsymbol{\theta}_g)$ to a convolutional neural network (CNN), \fedcombat could be adapted to handle image data directly, which would create new possibilities for harmonization in multi-centric imaging studies in federated setups. These considerations will potentially enhance the utility of \fedcombat, extending its reach in federated learning and data harmonization within healthcare research.

Simultaneously to this work, the recent work of \cite{gebre2023cross} illustrates the difficulties that harmonization methods face when dealing with imaging data, especially longitudinal data with scanner variations. Their study found that statistical methods like ComBat could match the distributions of cortical thickness across scanners but could not increase intra-class correlation coefficients (ICCs) or eliminate scanner change effects in longitudinal data sets. Deep learning methods like CycleGAN and Neural Style Transfer (NST) performed slightly better, but no method could completely harmonize the longitudinal dataset, underlining the complexity and difficulty of this problem. This suggests the necessity for more sophisticated methods capable of addressing these challenges centralized and inherently, federated scenarios. A recent deep learning-based \combat harmonization method (\deepcombat) was proposed by \cite{hu2024deepcombat}, using a conditional variational autoencoder (CVAE) to model both nonlinear and multivariate relation between covariates and cortical thickness measurements in a centralzed setting. Our proposed federated framework provides a generic optimization scheme that can be adapted in the future to extend this methodology to a decentralized setting.

\section{Ethics statement}
This study analyzes publicly available datasets.  All data collection protocols from the original studies were approved by their respective institutional review boards (IRB).

\section{Data and Code Availabilty}

The simulation code and the implementation of the proposed algorithms are available at: \url{https://gitlab.inria.fr/greguig/fedcombat}.

\section{Author Contributions}
S.S., G.R. and M.L. implemented the methods and conducted the experiments, while N.P.O. and A.A. contributed to the preprocessing of the data. S.S., N.P.O.,  A.A. and M.L. contributed to devising the core idea of the approach. N.P.O.,  A.A., and M.L. contributed to the revision of the manuscript. M.L. and A.A. contributed to the funding of the project. All authors read and approved the manuscript.

\section{Funding}
This project has received funding from the European Union's Horizon 2020 research and innovation programme under grant agreement No 847579 (Marie Skłodowska-Curie Actions), and by the INRIA Sophia Antipolis-Méditerranée "NEF" computation cluster. This work was partially supported by the Early Detection of Alzheimer’s Disease Subtypes (E-DADS) project, an EU Joint Programme - Neurodegenerative Disease Research (JPND) project (see www.jpnd.eu). The project is supported under the aegis of JPND through the following funding organizations: United Kingdom, Medical Research Council (MR/T046422/1); Netherlands, ZonMW (733051106); France, Agence Nationale de la Recherche (ANR-19-JPW2-000); Italy, Italian Ministry of Health (MoH); Australia, National Health \& Medical Research Council (1191535); Hungary, National Research, Development and Innovation Office (2019-2.1.7-ERA-NET-2020-00008). 

\section{Declaration of Competing Interests}
The authors declare no competing interests.

\section{Acknowledgments}
Data used in preparation of this article were obtained from the Alzheimer's Disease Neuroimaging Initiative (ADNI) database (\url{https://adni.loni.usc.edu/}). The investigators within the ADNI contributed to the design and implementation of ADNI and/or provided data but did not participate in analysis or writing of this report. A complete listing of ADNI investigators can be found at \url{https://adni.loni.usc.edu/wp-content/uploads/how_to_apply/ADNI_Acknowledgement_List.pdf}. Data collection and sharing for the Alzheimer's Disease Neuroimaging Initiative (ADNI) is funded by the National Institute on Aging (National Institutes of Health Grant U19AG024904). The grantee organization is the Northern California Institute for Research
and Education. In the past, ADNI has also received funding from the National Institute of Biomedical Imaging and Bioengineering, the Canadian Institutes of Health Research, and private sector contributions through the Foundation for the National Institutes of Health (FNIH) including generous contributions from the following: AbbVie, Alzheimer’s Association; Alzheimer’s Drug Discovery Foundation; Araclon Biotech; BioClinica, Inc.; Biogen; BristolMyers Squibb Company; CereSpir, Inc.; Cogstate; Eisai Inc.; Elan Pharmaceuticals, Inc.; Eli Lilly and Company; EuroImmun; F. Hoffmann-La Roche Ltd and its affiliated company Genentech, Inc.; Fujirebio; GE Healthcare; IXICO Ltd.; Janssen Alzheimer Immunotherapy
Research \& Development, LLC.; Johnson \& Johnson Pharmaceutical Research \& Development LLC.; Lumosity; Lundbeck; Merck \& Co., Inc.; Meso Scale Diagnostics, LLC.; NeuroRx Research; Neurotrack Technologies; Novartis Pharmaceuticals Corporation; Pfizer Inc.; Piramal Imaging; Servier; Takeda Pharmaceutical Company; and Transition Therapeutics.
\\ Data used in the preparation of this article was obtained from the Australian Imaging Biomarkers and Lifestyle flagship study of ageing (AIBL) funded by the Commonwealth Scientific and Industrial Research Organisation (CSIRO) which was made available at the ADNI database (\url{https://adni.loni.usc.edu/aibl-australian-imaging-biomarkers-and-lifestyle-study-of-ageing-18-month-data-now-released/}). Data used in the preparation of this article were obtained from the Autism Brain Imaging Data Exchange (ABIDE) I database. The AIBL researchers contributed data but did not participate in the analysis or writing of this report. AIBL researchers are listed at \url{https://www.aibl.csiro.au}. Data used in the preparation of this article were obtained from the Open Access Series of Imaging Studies (OASIS) database. Data used in the preparation of this article were obtained from the Parkinson's Progression Markers Initiative (PPMI) database (\url{http://www.ppmi-info.org/}). The A4 Study was a secondary prevention trial in preclinical Alzheimer's disease, aiming to slow cognitive decline associated with brain amyloid accumulation in clinically normal older individuals. The A4 Study was funded by a public-private-philanthropic partnership, including funding from the National Institutes of Health-National Institute on Aging, Eli Lilly and Company, Alzheimer's Association, Accelerating Medicines Partnership, GHR Foundation, an anonymous foundation, and additional private donors, with in-kind support from Avid Radiopharmaceuticals, Cogstate, Albert Einstein College of Medicine and the Foundation for Neurologic Diseases. The companion observational Longitudinal Evaluation of Amyloid Risk and Neurodegeneration (LEARN) Study was funded by the Alzheimer's Association and GHR Foundation. The A4 and LEARN Studies were led by Dr. Reisa Sperling at Brigham and Women's Hospital, Harvard Medical School, and Dr. Paul Aisen at the Alzheimer's Therapeutic Research Institute (ATRI) at the University of Southern California. The A4 and LEARN Studies were coordinated by ATRI at the University of Southern California, and the data are made available under the auspices of Alzheimer’s Clinical Trial Consortium through the Global Research \& Imaging Platform (GRIP). The complete A4 Study Team list is available on: \url{https://www.actcinfo.org/a4-study-team-lists/}. We would like to acknowledge the dedication of the study participants and their study partners who made the A4 and LEARN Studies possible.

\bibliographystyle{elsarticle-harv}

\begin{thebibliography}{23}
\expandafter\ifx\csname natexlab\endcsname\relax\def\natexlab#1{#1}\fi
\providecommand{\url}[1]{\texttt{#1}}
\providecommand{\href}[2]{#2}
\providecommand{\path}[1]{#1}
\providecommand{\DOIprefix}{doi:}
\providecommand{\ArXivprefix}{arXiv:}
\providecommand{\URLprefix}{URL: }
\providecommand{\Pubmedprefix}{pmid:}
\providecommand{\doi}[1]{\href{http://dx.doi.org/#1}{\path{#1}}}
\providecommand{\Pubmed}[1]{\href{pmid:#1}{\path{#1}}}
\providecommand{\bibinfo}[2]{#2}
\ifx\xfnm\relax \def\xfnm[#1]{\unskip,\space#1}\fi
\bibitem[{Ashburner and Friston(2005)}]{ashburner2005unified}
\bibinfo{author}{Ashburner, J.}, \bibinfo{author}{Friston, K.J.},
  \bibinfo{year}{2005}.
\newblock \bibinfo{title}{Unified segmentation}.
\newblock \bibinfo{journal}{Neuroimage} \bibinfo{volume}{26},
  \bibinfo{pages}{839--851}.
\bibitem[{Bethlehem et~al.(2022)Bethlehem, Seidlitz, White, Vogel, Anderson,
  Adamson, Adler, Alexopoulos, Anagnostou, Areces-Gonzalez
  et~al.}]{bethlehem2022brainNature}
\bibinfo{author}{Bethlehem, R.A.}, \bibinfo{author}{Seidlitz, J.},
  \bibinfo{author}{White, S.R.}, \bibinfo{author}{Vogel, J.W.},
  \bibinfo{author}{Anderson, K.M.}, \bibinfo{author}{Adamson, C.},
  \bibinfo{author}{Adler, S.}, \bibinfo{author}{Alexopoulos, G.S.},
  \bibinfo{author}{Anagnostou, E.}, \bibinfo{author}{Areces-Gonzalez, A.},
  et~al., \bibinfo{year}{2022}.
\newblock \bibinfo{title}{Brain charts for the human lifespan}.
\newblock \bibinfo{journal}{Nature} \bibinfo{volume}{604},
  \bibinfo{pages}{525--533}.
\bibitem[{Chen et~al.(2022)Chen, Luo, Chen, Shinohara, Shou, Initiative
  et~al.}]{dcombat}
\bibinfo{author}{Chen, A.A.}, \bibinfo{author}{Luo, C.}, \bibinfo{author}{Chen,
  Y.}, \bibinfo{author}{Shinohara, R.T.}, \bibinfo{author}{Shou, H.},
  \bibinfo{author}{Initiative, A.D.N.}, et~al., \bibinfo{year}{2022}.
\newblock \bibinfo{title}{Privacy-preserving harmonization via distributed
  combat}.
\newblock \bibinfo{journal}{Neuroimage} \bibinfo{volume}{248},
  \bibinfo{pages}{118822}.
\bibitem[{Coup{\'e} et~al.(2017)Coup{\'e}, Catheline, Lanuza, Manj{\'o}n and
  Initiative}]{coupe2017towards}
\bibinfo{author}{Coup{\'e}, P.}, \bibinfo{author}{Catheline, G.},
  \bibinfo{author}{Lanuza, E.}, \bibinfo{author}{Manj{\'o}n, J.V.},
  \bibinfo{author}{Initiative, A.D.N.}, \bibinfo{year}{2017}.
\newblock \bibinfo{title}{Towards a unified analysis of brain maturation and
  aging across the entire lifespan: A mri analysis}.
\newblock \bibinfo{journal}{Human brain mapping} \bibinfo{volume}{38},
  \bibinfo{pages}{5501--5518}.
\bibitem[{Dale et~al.(1999)Dale, Fischl and Sereno}]{dale1999cortical}
\bibinfo{author}{Dale, A.M.}, \bibinfo{author}{Fischl, B.},
  \bibinfo{author}{Sereno, M.I.}, \bibinfo{year}{1999}.
\newblock \bibinfo{title}{Cortical surface-based analysis: I. segmentation and
  surface reconstruction}.
\newblock \bibinfo{journal}{Neuroimage} \bibinfo{volume}{9},
  \bibinfo{pages}{179--194}.
\bibitem[{Desikan et~al.(2006)Desikan, S{\'e}gonne, Fischl, Quinn, Dickerson,
  Blacker, Buckner, Dale, Maguire, Hyman et~al.}]{desikan2006automated}
\bibinfo{author}{Desikan, R.S.}, \bibinfo{author}{S{\'e}gonne, F.},
  \bibinfo{author}{Fischl, B.}, \bibinfo{author}{Quinn, B.T.},
  \bibinfo{author}{Dickerson, B.C.}, \bibinfo{author}{Blacker, D.},
  \bibinfo{author}{Buckner, R.L.}, \bibinfo{author}{Dale, A.M.},
  \bibinfo{author}{Maguire, R.P.}, \bibinfo{author}{Hyman, B.T.}, et~al.,
  \bibinfo{year}{2006}.
\newblock \bibinfo{title}{An automated labeling system for subdividing the
  human cerebral cortex on mri scans into gyral based regions of interest}.
\newblock \bibinfo{journal}{Neuroimage} \bibinfo{volume}{31},
  \bibinfo{pages}{968--980}.
\bibitem[{Ducharme et~al.(2016)Ducharme, Albaugh, Nguyen, Hudziak,
  Mateos-P{\'e}rez, Labbe, Evans, Karama, Group
  et~al.}]{ducharme2016trajectories}
\bibinfo{author}{Ducharme, S.}, \bibinfo{author}{Albaugh, M.D.},
  \bibinfo{author}{Nguyen, T.V.}, \bibinfo{author}{Hudziak, J.J.},
  \bibinfo{author}{Mateos-P{\'e}rez, J.M.}, \bibinfo{author}{Labbe, A.},
  \bibinfo{author}{Evans, A.C.}, \bibinfo{author}{Karama, S.},
  \bibinfo{author}{Group, B.D.C.}, et~al., \bibinfo{year}{2016}.
\newblock \bibinfo{title}{Trajectories of cortical thickness maturation in
  normal brain development—the importance of quality control procedures}.
\newblock \bibinfo{journal}{Neuroimage} \bibinfo{volume}{125},
  \bibinfo{pages}{267--279}.
\bibitem[{Fischl et~al.(2004)Fischl, Van Der~Kouwe, Destrieux, Halgren,
  S{\'e}gonne, Salat, Busa, Seidman, Goldstein, Kennedy
  et~al.}]{fischl2004automatically}
\bibinfo{author}{Fischl, B.}, \bibinfo{author}{Van Der~Kouwe, A.},
  \bibinfo{author}{Destrieux, C.}, \bibinfo{author}{Halgren, E.},
  \bibinfo{author}{S{\'e}gonne, F.}, \bibinfo{author}{Salat, D.H.},
  \bibinfo{author}{Busa, E.}, \bibinfo{author}{Seidman, L.J.},
  \bibinfo{author}{Goldstein, J.}, \bibinfo{author}{Kennedy, D.}, et~al.,
  \bibinfo{year}{2004}.
\newblock \bibinfo{title}{Automatically parcellating the human cerebral
  cortex}.
\newblock \bibinfo{journal}{Cerebral cortex} \bibinfo{volume}{14},
  \bibinfo{pages}{11--22}.
\bibitem[{Fortin et~al.(2018)Fortin, Cullen, Sheline, Taylor, Aselcioglu, Cook,
  Adams, Cooper, Fava, McGrath et~al.}]{neurocombat}
\bibinfo{author}{Fortin, J.P.}, \bibinfo{author}{Cullen, N.},
  \bibinfo{author}{Sheline, Y.I.}, \bibinfo{author}{Taylor, W.D.},
  \bibinfo{author}{Aselcioglu, I.}, \bibinfo{author}{Cook, P.A.},
  \bibinfo{author}{Adams, P.}, \bibinfo{author}{Cooper, C.},
  \bibinfo{author}{Fava, M.}, \bibinfo{author}{McGrath, P.J.}, et~al.,
  \bibinfo{year}{2018}.
\newblock \bibinfo{title}{Harmonization of cortical thickness measurements
  across scanners and sites}.
\newblock \bibinfo{journal}{Neuroimage} \bibinfo{volume}{167},
  \bibinfo{pages}{104--120}.
\bibitem[{Fortin et~al.(2017)Fortin, Parker, Tun{\c{c}}, Watanabe, Elliott,
  Ruparel, Roalf, Satterthwaite, Gur, Gur et~al.}]{fortin2017harmonization}
\bibinfo{author}{Fortin, J.P.}, \bibinfo{author}{Parker, D.},
  \bibinfo{author}{Tun{\c{c}}, B.}, \bibinfo{author}{Watanabe, T.},
  \bibinfo{author}{Elliott, M.A.}, \bibinfo{author}{Ruparel, K.},
  \bibinfo{author}{Roalf, D.R.}, \bibinfo{author}{Satterthwaite, T.D.},
  \bibinfo{author}{Gur, R.C.}, \bibinfo{author}{Gur, R.E.}, et~al.,
  \bibinfo{year}{2017}.
\newblock \bibinfo{title}{Harmonization of multi-site diffusion tensor imaging
  data}.
\newblock \bibinfo{journal}{Neuroimage} \bibinfo{volume}{161},
  \bibinfo{pages}{149--170}.
\bibitem[{Gebre et~al.(2023)Gebre, Senjem, Raghavan, Schwarz, Gunter,
  Hofrenning, Reid, Kantarci, Graff-Radford, Knopman et~al.}]{gebre2023cross}
\bibinfo{author}{Gebre, R.K.}, \bibinfo{author}{Senjem, M.L.},
  \bibinfo{author}{Raghavan, S.}, \bibinfo{author}{Schwarz, C.G.},
  \bibinfo{author}{Gunter, J.L.}, \bibinfo{author}{Hofrenning, E.I.},
  \bibinfo{author}{Reid, R.I.}, \bibinfo{author}{Kantarci, K.},
  \bibinfo{author}{Graff-Radford, J.}, \bibinfo{author}{Knopman, D.S.}, et~al.,
  \bibinfo{year}{2023}.
\newblock \bibinfo{title}{Cross--scanner harmonization methods for structural
  mri may need further work: A comparison study}.
\newblock \bibinfo{journal}{NeuroImage} \bibinfo{volume}{269},
  \bibinfo{pages}{119912}.
\bibitem[{Hu et~al.(2024)Hu, Lucas, Chen, Coleman, Horng, Ng, Tustison, Davis,
  Shou, Li et~al.}]{hu2024deepcombat}
\bibinfo{author}{Hu, F.}, \bibinfo{author}{Lucas, A.}, \bibinfo{author}{Chen,
  A.A.}, \bibinfo{author}{Coleman, K.}, \bibinfo{author}{Horng, H.},
  \bibinfo{author}{Ng, R.W.}, \bibinfo{author}{Tustison, N.J.},
  \bibinfo{author}{Davis, K.A.}, \bibinfo{author}{Shou, H.},
  \bibinfo{author}{Li, M.}, et~al., \bibinfo{year}{2024}.
\newblock \bibinfo{title}{Deepcombat: A statistically motivated,
  hyperparameter-robust, deep learning approach to harmonization of
  neuroimaging data}.
\newblock \bibinfo{journal}{Human brain mapping} \bibinfo{volume}{45},
  \bibinfo{pages}{e26708}.
\bibitem[{Johnson et~al.(2007)Johnson, Li and Rabinovic}]{combat}
\bibinfo{author}{Johnson, W.E.}, \bibinfo{author}{Li, C.},
  \bibinfo{author}{Rabinovic, A.}, \bibinfo{year}{2007}.
\newblock \bibinfo{title}{Adjusting batch effects in microarray expression data
  using empirical bayes methods}.
\newblock \bibinfo{journal}{Biostatistics} \bibinfo{volume}{8},
  \bibinfo{pages}{118--127}.
\bibitem[{Kone{\v{c}}n{\`y} et~al.(2016)Kone{\v{c}}n{\`y}, McMahan, Yu,
  Richt{\'a}rik, Suresh and Bacon}]{konevcny2016federated}
\bibinfo{author}{Kone{\v{c}}n{\`y}, J.}, \bibinfo{author}{McMahan, H.B.},
  \bibinfo{author}{Yu, F.X.}, \bibinfo{author}{Richt{\'a}rik, P.},
  \bibinfo{author}{Suresh, A.T.}, \bibinfo{author}{Bacon, D.},
  \bibinfo{year}{2016}.
\newblock \bibinfo{title}{Federated learning: Strategies for improving
  communication efficiency}.
\newblock \bibinfo{journal}{arXiv preprint arXiv:1610.05492} .
\bibitem[{Li et~al.(2020)Li, Sahu, Zaheer, Sanjabi, Talwalkar and
  Smith}]{li2020federated}
\bibinfo{author}{Li, T.}, \bibinfo{author}{Sahu, A.K.},
  \bibinfo{author}{Zaheer, M.}, \bibinfo{author}{Sanjabi, M.},
  \bibinfo{author}{Talwalkar, A.}, \bibinfo{author}{Smith, V.},
  \bibinfo{year}{2020}.
\newblock \bibinfo{title}{Federated optimization in heterogeneous networks}.
\newblock \bibinfo{journal}{Proceedings of Machine learning and systems}
  \bibinfo{volume}{2}, \bibinfo{pages}{429--450}.
\bibitem[{Li et~al.(2019)Li, Huang, Yang, Wang and Zhang}]{li2019convergence}
\bibinfo{author}{Li, X.}, \bibinfo{author}{Huang, K.}, \bibinfo{author}{Yang,
  W.}, \bibinfo{author}{Wang, S.}, \bibinfo{author}{Zhang, Z.},
  \bibinfo{year}{2019}.
\newblock \bibinfo{title}{On the convergence of fedavg on non-iid data}.
\newblock \bibinfo{journal}{arXiv preprint arXiv:1907.02189} .
\bibitem[{McMahan et~al.(2017)McMahan, Moore, Ramage, Hampson and
  y~Arcas}]{fedavg}
\bibinfo{author}{McMahan, B.}, \bibinfo{author}{Moore, E.},
  \bibinfo{author}{Ramage, D.}, \bibinfo{author}{Hampson, S.},
  \bibinfo{author}{y~Arcas, B.A.}, \bibinfo{year}{2017}.
\newblock \bibinfo{title}{Communication-efficient learning of deep networks
  from decentralized data}, in: \bibinfo{booktitle}{Artificial intelligence and
  statistics}, \bibinfo{organization}{PMLR}. pp. \bibinfo{pages}{1273--1282}.
\bibitem[{Plis et~al.(2016)Plis, Sarwate, Wood, Dieringer, Landis, Reed, Panta,
  Turner, Shoemaker, Carter et~al.}]{plis2016coinstac}
\bibinfo{author}{Plis, S.M.}, \bibinfo{author}{Sarwate, A.D.},
  \bibinfo{author}{Wood, D.}, \bibinfo{author}{Dieringer, C.},
  \bibinfo{author}{Landis, D.}, \bibinfo{author}{Reed, C.},
  \bibinfo{author}{Panta, S.R.}, \bibinfo{author}{Turner, J.A.},
  \bibinfo{author}{Shoemaker, J.M.}, \bibinfo{author}{Carter, K.W.}, et~al.,
  \bibinfo{year}{2016}.
\newblock \bibinfo{title}{Coinstac: a privacy enabled model and prototype for
  leveraging and processing decentralized brain imaging data}.
\newblock \bibinfo{journal}{Frontiers in neuroscience} \bibinfo{volume}{10},
  \bibinfo{pages}{365}.
\bibitem[{Pomponio et~al.(2020)Pomponio, Erus, Habes, Doshi, Srinivasan,
  Mamourian, Bashyam, Nasrallah, Satterthwaite, Fan et~al.}]{combatgam}
\bibinfo{author}{Pomponio, R.}, \bibinfo{author}{Erus, G.},
  \bibinfo{author}{Habes, M.}, \bibinfo{author}{Doshi, J.},
  \bibinfo{author}{Srinivasan, D.}, \bibinfo{author}{Mamourian, E.},
  \bibinfo{author}{Bashyam, V.}, \bibinfo{author}{Nasrallah, I.M.},
  \bibinfo{author}{Satterthwaite, T.D.}, \bibinfo{author}{Fan, Y.}, et~al.,
  \bibinfo{year}{2020}.
\newblock \bibinfo{title}{Harmonization of large mri datasets for the analysis
  of brain imaging patterns throughout the lifespan}.
\newblock \bibinfo{journal}{NeuroImage} \bibinfo{volume}{208},
  \bibinfo{pages}{116450}.
\bibitem[{Reynolds et~al.(2022)Reynolds, Chaudhary, Torbati, Tudorascu and
  Batmanghelich}]{reynolds2022combat}
\bibinfo{author}{Reynolds, M.}, \bibinfo{author}{Chaudhary, T.},
  \bibinfo{author}{Torbati, M.E.}, \bibinfo{author}{Tudorascu, D.L.},
  \bibinfo{author}{Batmanghelich, K.}, \bibinfo{year}{2022}.
\newblock \bibinfo{title}{Combat harmonization: Empirical bayes versus fully
  bayes approaches}.
\newblock \bibinfo{journal}{bioRxiv} .
\bibitem[{Silva et~al.(2020)Silva, Altmann, Gutman and Lorenzi}]{silva2020fed}
\bibinfo{author}{Silva, S.}, \bibinfo{author}{Altmann, A.},
  \bibinfo{author}{Gutman, B.}, \bibinfo{author}{Lorenzi, M.},
  \bibinfo{year}{2020}.
\newblock \bibinfo{title}{Fed-biomed: A general open-source frontend framework
  for federated learning in healthcare}, in: \bibinfo{booktitle}{Domain
  Adaptation and Representation Transfer, and Distributed and Collaborative
  Learning: Second MICCAI Workshop, DART 2020, and First MICCAI Workshop, DCL
  2020, Held in Conjunction with MICCAI 2020, Lima, Peru, October 4--8, 2020,
  Proceedings 2}, \bibinfo{organization}{Springer}. pp.
  \bibinfo{pages}{201--210}.
\bibitem[{Sled et~al.(1998)Sled, Zijdenbos and Evans}]{sled1998nonparametric}
\bibinfo{author}{Sled, J.G.}, \bibinfo{author}{Zijdenbos, A.P.},
  \bibinfo{author}{Evans, A.C.}, \bibinfo{year}{1998}.
\newblock \bibinfo{title}{A nonparametric method for automatic correction of
  intensity nonuniformity in mri data}.
\newblock \bibinfo{journal}{IEEE transactions on medical imaging}
  \bibinfo{volume}{17}, \bibinfo{pages}{87--97}.
\bibitem[{Wachinger et~al.(2021)Wachinger, Rieckmann, P{\"o}lsterl, Initiative
  et~al.}]{wachinger2021detect}
\bibinfo{author}{Wachinger, C.}, \bibinfo{author}{Rieckmann, A.},
  \bibinfo{author}{P{\"o}lsterl, S.}, \bibinfo{author}{Initiative, A.D.N.},
  et~al., \bibinfo{year}{2021}.
\newblock \bibinfo{title}{Detect and correct bias in multi-site neuroimaging
  datasets}.
\newblock \bibinfo{journal}{Medical Image Analysis} \bibinfo{volume}{67},
  \bibinfo{pages}{101879}.

\end{thebibliography}

\end{document}